# Assessing the impact of Introductory Physics for the Life Sciences on students' ability to build complex models

Benjamin D. Geller,* Maya Tipton, Brandon Daniel-Morales,† Nikhil Tignor,† Calvin White,† and Catherine H. Crouch

*Department of Physics & Astronomy, Swarthmore College, Swarthmore, PA 19081*

## Abstract

A central goal of Introductory Physics for the Life Sciences (IPLS) is to prepare students to use physics to model and analyze biological situations, a skill of increasing importance for their future studies and careers. Here we report our findings on life science students' ability to carry out a sophisticated biological modeling task at the end of first-semester introductory physics. Some students were enrolled in a standard course ($N = 34$), and some in an IPLS course ($N = 61$); both courses were taught with active learning, used calculus, and included the same core physics concepts. Compared to those who took the standard course, we found that the IPLS students were significantly more successful at building a model that combined ideas in a manner they had not previously seen, and at making complex decisions about how to apply an equation to a particular physical situation, although both groups displayed similar success at solving simpler problems. Both groups identified and applied simple models that they had previously used in very similar contexts, and executed straightforward calculations, at statistically indistinguishable rates. We report both our findings and the rationale behind the development of the task, in the hopes that others may find this task either a valuable tool or a starting point to develop other such tasks. Further study is needed to determine the basis for the IPLS students' stronger performance—namely, what aspects of the IPLS course supports these students to be better prepared to do such modeling—as well as whether biological settings are important for IPLS students to succeed in flexible model building, and whether the ability to employ models flexibly persists over time.

* Corresponding author. † These authors contributed similarly to this work and are listed alphabetically.

## I. INTRODUCTION: PREPARING LIFE SCIENCE STUDENTS TO USE PHYSICAL MODELS FLEXIBLY

Over the last decade, several investigators have developed and begun assessing introductory physics courses specifically aimed at supporting undergraduate life science and pre-medical students to gain a deeper understanding of the physical sciences, and to apply this understanding to the life and medical sciences [1–6]. These Introductory Physics for the Life Sciences (IPLS) courses were developed in response to calls from national professional societies in the life sciences and medicine, and seek to support students in developing problem solving and mathematical skills as well as topical understanding [7–10]. A common feature of such courses is a central and repeated emphasis on analyzing complex biological situations using simple physical models that are nonetheless mathematically accessible within an introductory calculus-based physics course.

As life science students move on to their future studies and careers, they will be required to implement physical models in a *flexible* way. That is, they will encounter complex biological problems that require them to choose between models in ways that are appropriate for the particular situation being modeled [8,10]. In many cases these situations will be more complex than those encountered in introductory physics. In the work reported here, we seek to assess life science students' ability to use physical models in this flexible way. We analyze life science students' work on a task given at the end of the mechanics semester in both IPLS and standard introductory physics. In this task, students analyzed an unfamiliar biological situation involving fluid dynamics, a situation for which successful analysis required combining model elements in a manner not previously demonstrated in either course. The goal in assessing student work at the end of the mechanics semester was to establish whether IPLS prepares life science students to engage in the sort of complex modeling that will be required of them in their future studies and careers.

We found that IPLS students displayed a significantly greater ability to develop a complex model combining two different physical concepts, although both IPLS and non-IPLS students displayed similar ability to apply familiar models, both in that same task and in a companion problem that did not require the same amount of modeling. This suggests that IPLS students may be better prepared to independently and spontaneously apply physics to biological situations in which it is needed, consistent with preliminary evidence from our work analyzing embedded tasks in intermediate biology courses [11] and from our longitudinal study of IPLS student work in a biology capstone course [12].

In this article we report both our findings and the thinking behind the task design, in the hopes that others may find this task a valuable tool or a starting point to develop their own. Section II briefly describes the theoretical framework underlying the design of the IPLS course at Swarthmore; Section III describes the two physics courses from which the study subjects were drawn; Section IV describes the task design and Section V its administration and analysis; Section VI presents results and our interpretation; and Section VII offers open questions and next steps.

## II. PREPARATION FOR FUTURE LEARNING: A THEORETICAL FRAMEWORK FOR IPLS AT SWARTHMORE

The Swarthmore IPLS course is designed to support students in using the tools and concepts of physics in their later biomedical studies and careers, and thus it ultimately has interdisciplinary transfer as a core goal [5]. This goal is grounded in the "Preparation for Future Learning" (PFL) concept of transfer [13–15]. We are not seeking to prepare students to be expert biological physicists in a single course, which seems unrealistic. Rather, our goal is to prepare students to recognize biological and biomedical situations in which using physical models will add to their understanding, and to make them more amenable to seeking the resources needed to do so. When attempting to analyze new biological contexts in their later work, we hope that our students will be prepared to employ the physical models that they encountered in IPLS, including possibly seeking additional resources to help them do so, and to apply these models flexibly. Indeed, the primary focus when analyzing student work on the fluid dynamics task described in this paper was to assess the degree to which IPLS and non-IPLS students were successful at doing just this.

In the IPLS course at Swarthmore, the instructor regularly demonstrates the process of complex problem-solving as part of an interactive lecture, with particular attention given to the decision-making steps and simplifying assumptions that are essential for describing complex biological systems with simple physical models. The instructor regularly describes the limitations of a particular model as it is being employed, and why a more complicated situation might necessitate that multiple models be combined. This explicit and repeated emphasis on the modeling process across a wide range of biological contexts is designed to prepare students to use simple physical models in a flexible way, as opposed to rigidly associating a single model with a single biological scenario. In Section III we will describe some specific conceptual areas in which students were exposed to these modeling choices during the IPLS course.

In addition to a recurring and explicit emphasis on the modeling process, the Swarthmore IPLS course employs two pedagogical strategies that have been shown to support a PFL mode of transfer: expansive framing [16] and cognitive apprenticeship [17]. Expansive framing refers to a curricular presentation that allows students to see the course topics as broadly applicable to the scientific community outside the physics classroom, and to their own interests and future careers [16]. The course content is explicitly presented ("framed") as being relevant and connected to students' other coursework in biology and chemistry, both now and in the future. Such framing has been shown to support life science student interest and engagement in physics [18,19], and this interest and engagement may in turn be important for the development of problem-solving skill and persistence [20].

In the cognitive apprenticeship framework, the goal is to create a learning environment that has essential features in common with the environment in which an expert functions [17]. Specifically, such an environment repeatedly prompts the apprentice to assess (i) why they are learning what they are learning, and (ii) how what they are learning connects to things they already know. These metacognitive practices are encouraged and assessed in the Swarthmore IPLS course, and are essential for the sort of flexible model implementation that is being assessed in this study.

Although the Swarthmore IPLS course is intended to foster a PFL mode of transfer, the study described here seeks only to evaluate students' ability to use models in a flexible way. Due to our constraints in delivering the assessment, and our more limited goals, this study does not incorporate the full design used in seminal studies of PFL [13,15]. In particular, students are instructed to use only the resources with which we provide them and with which they are familiar from the introductory mechanics context. In our long-term goals for our students, congruent with the PFL framework, we expect students will be able to seek out and use appropriate resources to solve new problems, but we do not assess that in this study. As described in Section IV, some parts of the task are more closely aligned with prior studies of PFL than others. Nevertheless, the task as a whole assesses students' ability to employ and combine models in ways that will be required of them in their future learning.

## III. STUDENT GROUPS AND COURSE CONTEXTS

In this study, we compared the modeling skills of life science and pre-medical students from different introductory physics courses by analyzing written work from 95 life science and pre-medical students enrolled in either IPLS ($N = 61$) or standard ($N = 34$) introductory mechanics.

### A. Subjects

Although life science and pre-medical students at Swarthmore typically take the IPLS course, during Year 1 of this study, due to staffing limitations, only the standard mechanics course was offered. For this reason, 29 of the 34 non-IPLS students in our study took the standard course during Year 1; the remaining 5 non-IPLS students in our study took the standard course in Year 2, despite IPLS being offered, most likely due to a schedule conflict. All 61 IPLS students took the IPLS course in Year 2.

We expected that other than taking different introductory physics courses, the IPLS and non-IPLS students came from equivalent populations, as they represented all of the life science and pre-medical students enrolled in first semester introductory physics in a given year. We found that the distribution of the students in the two groups was indistinguishable across class years (first-years, sophomores, juniors, seniors; Fisher's exact test, $p = 0.436$) and majors (biology/neuroscience, chemistry/biochemistry, or pre-medical students with more quantitative majors such as engineering or mathematics; Fisher's exact test, $p = 0.942$).

### B. Courses

The same instructor taught the standard course both years; no IPLS course was offered in Year 1 and BDG taught the IPLS course in Year 2. During Year 1, the standard course was the only first semester mechanics course for non-physics majors offered. It enrolled 81 students, meeting in two sections. (This course included engineering as well as life science students; although we collected responses from the engineering students, we do not analyze them here.) During Year 2, the IPLS course enrolled 66 students and the non-IPLS course enrolled 47 students (mostly engineering students), each taught in a single section.

Both courses used active learning strategies in class, assigned students to read before class and complete pre-class questionnaires (modeled on Just-in-time Teaching PreFlights [21]), and used the same textbook (Knight, *Physics for Scientists and Engineers*, 3rd edition). Both BDG and the standard course instructor had several previous years of experience teaching with active learning methods, although the standard course instructor was new to Swarthmore during Year 1 while BDG had taught at Swarthmore for 3 years.

Both courses covered the same topics in fluid statics and dynamics, including both the Bernoulli equation and the Hagen-Poiseuille relationship for viscous flow. Both courses also required a mix of conceptual reasoning and quantitative problem solving, although the IPLS course required somewhat less mathematically involved problems. Both were calculus-based, although the standard course used calculus somewhat more extensively.

Both courses articulated many similar learning goals for students in the syllabus. The standard course named its objectives as follows:

> Proficiency in physics includes not only the particular knowledge of the subject, but also skills in using such knowledge. To name just a few:
>
> - ***Sensemaking.*** The ability to "make sense" out of observed phenomena by applying theories and models in physics qualitatively.
> - ***Problem solving***. The ability to formulate problems concretely, visualize the situations in physical terms, and plan and execute the appropriate solution strategies—rather than just scrambling a bunch of equations together.
> - ***Estimation***. The ability to get a sense of how big/small the answers of a prob- lem should be without involved calculations—not all problems require precise answers, and even those that do can make use of some checks.
> - ***Metacognition***, which means thinking about thinking. In other words, the aware- ness of what you know, what you don't know, what you need to know, etc.
> - ***Communication and collaboration***. Contrary to popular beliefs physicists do work together—some of our problems are simply too big to tackle alone. As such, it is important to learn how to talk and work physics with each other.

And the IPLS course named its objectives as follows:

> Our focus over the next four months will be on those physical models and ideas that are especially relevant for understanding the living world:
>
> - developing a deep conceptual understanding of the fundamental principles of motion, force, and energy, one that you can clearly articulate both in words and with mathematics
> - relating this conceptual understanding to other ideas that you have already encountered in your biology, chemistry, and mathematics courses
> - developing both the qualitative reasoning skills (metacognition, checking for order of magnitude reasonableness, etc.) and quantitative problem-solving skills (estimation, modeling, etc.) that will help you to apply the physics we learn to biological problems you may not yet have seen
>
> Most importantly, this course is designed to help you **identify and navigate the disciplinary boundaries** between biology, chemistry, and physics, and to **make your experience across these boundaries more coherent**.

How might these courses differently develop students' abilities to model biological situations? The IPLS course explicitly communicates that selecting or building simple physical models is an essential skill for analyzing biomedical phenomena with physics, and assigned tasks explicitly involving modeling choices. The standard class did not explicitly focus on or teach the modeling process, although the textbook described the use of physics in terms of models and the instructor occasionally used that language in class. In addition, each unit of the Swarthmore IPLS course is built around *authentic* biological contexts [22,23], drawn directly from examples that students encounter in their life science classes, which we have previously demonstrated leads to students' increased interest and sense of the relevance of physics [19]. In the fluid dynamics unit that is central to this study, for example, the principles of fluid dynamics were motivated as essential for understanding normal and abnormal human cardiology (Appendix A). The biological contexts in the IPLS course were integral to the course and repeatedly referred to throughout each unit as the physical ideas were developed. The standard course offered a few biologically relevant problems as applications at the end of a unit, but the course was not organized around those problems.

Examples of the opportunities provided in the IPLS course for students to choose a simple model, and to identify the physical parameters that guide model choice, included:

- During the study of resistance to motion through fluids, students were introduced to models of the resistive force as proportional to the object's speed (linear) or the speed squared (quadratic), and assigned problems in which they had to choose an appropriate model based on information given in the problem. For example, students were provided with multiple scenarios in which objects of different sizes move through different fluids, and had to choose whether the linear or quadratic model was most appropriate for each.
- When learning about the elastic stretching of bones and ligaments, students were asked to carefully consider the limitations of Hooke's law; Hooke's law was framed as a model that must be applied flexibly, through careful consideration of the particular biophysical situation. Students were presented with multiple situations in which Hooke's law was an inadequate model, and were assigned a problems in which they had to determine from force-vs-extension data whether Hooke's Law was an appropriate model for the system.
- For several different types of biological motion, IPLS students were asked to determine whether directed or diffusive (random) motion would be faster for the length scales involved, and which was a better model for the particular motion being considered.

- In the fluid dynamics unit, IPLS students worked through a series of problems and questions related to the human cardiovascular system (Appendix A). As in previous units, they were explicitly asked to select the most appropriate fluid dynamics model. Students learned that when blood flows through a wide, short orifice, such as the aortic valve between the left ventricle and the left atrium, it is appropriate to model the flow using the Bernoulli equation; when blood flows through vessels that are relatively long compared to their diameter, the Hagen-Poiseuille model appropriately accounts for viscous resistance.

## IV. TASK DESIGN

The goal of this task was to learn whether IPLS students displayed a different level of success than their non-IPLS peers in (i) modeling unfamiliar biological situations using physics, and (ii) combining physics concepts in novel ways to model such situations. To assess this required a biologically relevant situation which could be analyzed with introductory physics, but which had not been used previously in the course, and for which a correct analysis required combining the physical concepts taught in both courses in a new manner. We devised a task that had all of these features, and which also had not been used in previous years of either course, so that students who received course materials from previous students would not have seen solutions to a related problem. The quality of this task was essential to the success of this project, so we describe the design process here in detail.

We designed a task analyzing the pressure difference between the roots and leaves of trees, an example which was not covered in either course, and which also was not discussed in Swarthmore's introductory biology course. The task was designed to evaluate whether students could develop a model they had not been taught, by combining the viscous flow model in the form of the Hagen-Poiseuille equation, which describes flow through horizontal pipes, with a term that modeled the effect of gravity on pressure. Previously, the effect of gravity on pressure had been introduced only in the context of fluid statics and non-viscous flow, so students were not shown how to include the effects of gravity when analyzing viscous flow. Furthermore, the problem statement did not explicitly invoke the ideas of either viscous or non-viscous flow, and did not explicitly mention any of the relevant equations or relationships.

The task also included a textbook thermodynamics problem, similar to those assigned in both courses, to provide an additional measure of students' general physics problem solving skills.

Equations and constants related to fluids and thermodynamics were provided on a reference sheet, as both courses provided such reference sheets for all tests.

The fluids problem appears part-by-part in Figures 1-3; the full task, with initial instructions, equations, and the thermodynamics problem, is provided in Appendix B, and a minimal solution in Appendix C. The remainder of this section presents the detailed design logic of the task.

### A. First part: Reminder of static pressure dependence on height, and measure of students' ability to apply a simple model

Part (a) of the task (Fig. 1), in which students found the height dependence of a giraffe's blood pressure, served two purposes: (i) to remind students about role of gravity in fluid pressure, and (ii) to determine their ability to apply a simple model learned in class to a new situation.

a) Adult male giraffes can reach a height of roughly 6 m. The minimum pressure of the blood leaving the giraffe's heart is 1.24 atmospheres (124 kPa). (Although blood is a mixture of water and various types of blood cells, the density of blood is very close to the density of water because the cells also consist mostly of water ) Find an approximate value for the minimum blood pressure in the giraffe's brain when its neck is extended to its full height. You may infer information from the picture of a giraffe provided.

***Please briefly explain the reasoning you used to find your answer, including how you decided which equations to use, as well any approximations you made. Also please show your work***

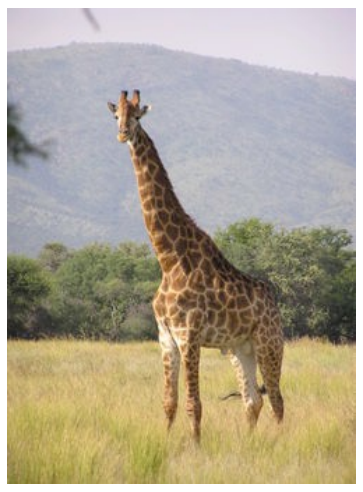

**FIG. 1.** Part (a) of the fluids problem. Photo from Wikimedia Commons, Miroslav Ducacheck, CC BY-SA 3.0. The full task (including instructions preceding the fluids problem, the thermodynamics problem, and the list of equations) is provided in Appendix B.

Detailed analysis (beyond the scope of the course) of the situation reveals that the dynamic corrections are small compared to the fluid static effects, so we felt it was reasonable to design this part so that students had enough information to analyze the fluid statics effects, and none of the information they would need to incorporate the dynamics. Some students did attempt to account for flow speed as well as gravity by using the Bernoulli equation, but all of them found ways (of variable correctness) to eliminate dependence on the (unspecified) flow speed, rather than estimating flow speeds. (The Bernoulli equation in the form presented in introductory physics does not account for the effects of branching, which is significant in the mammalian circulatory system, so using that form of the Bernoulli equation with physiologically accurate flow speeds gives a very misleading result.)

As students in both courses had done problems calculating the effect of fluid pressure on height, this part of the task also allowed us to evaluate students' level of problem-solving skill in a context very similar to problems previously encountered. The thermodynamics problem (section IV.D) also allowed us to assess baseline problem-solving skill.

### B. Second part: identifying nature of flow and incorporating gravity

Part (b) of the task (Fig. 2) provided a measure of students' ability to build new models using the physics they had learned. For a fully correct answer, students needed to combine the effect of gravity with the Hagen-Poiseuille equation for viscous flow through a horizontal cylindrical tube. The viscous nature of the flow had to be inferred from the dimensions of the tubes (xylem), as no explicit mention of viscosity was made. We expected this would be extremely challenging.

b) In trees, water is carried from the roots to the leaves by the flow of sap (water with other kinds of molecules dissolved in it) through stiff tube-like structures, called xylem. Although sizes vary, a typical diameter would be 100 μm. In the main trunk of the tree, they extend close to the full height of the tree, which is commonly as great as 30 meters tall or taller (5 species of tree are known to reach 90 -110 m in height). These extremely narrow, long tubes, called xylem, contain a continuous column of water which can then flow into the leaves. The evaporation of water from the leaves (called transpiration) causes water to be steadily drawn into the leaves from the xylem. The structure of the leaves allows the pressure of water in the xylem to not necessarily be the same as the surrounding atmospheric pressure.

Consider a tree in which sap flows through each 100 μm-diameter xylem at a volume flow rate of $1.1\text{x}10^{-10}$ $m^3$/s (equal to $1.1\text{x}10^{-4}$ mL/s or 0.40 mL/hr), corresponding to an average flow speed of 0.014 m/s. (Given the huge number of xylem, the total flow for the entire tree is substantial!) If the pressure in the roots is equal to atmospheric pressure, what is the pressure at the top of a 30 m tall xylem in the trunk?

***Please briefly explain the reasoning you used to find your answer, including how you decided which equations to use, as well any approximations you made. Also please show your work.***

**FIG. 2.** Part (b) of the fluids problem. The full task (with additional instructions before the problem, the thermodynamics problem, and the list of equations) is provided in Appendix B.

Students were told to find the pressure inside the trunk at the top so that they did not have to also account for surface tension of the water exiting the leaf pores, or worry about partial pressures and phases. They were also explicitly told that the pressure in the leaves did not have to match the surrounding atmospheric pressure. The tree height was chosen so that students would obtain a negative pressure at the top of the trunk if they correctly implemented any model involving gravity, whether using viscous or non-viscous flow. Flow speeds and xylem sizes were taken from Niklas and Spatz, *Plant Physics* [24] (Ch. 3 and particularly section 3.8 discuss "the ascent of water" in trees in advanced detail).

### C. Third part: considering the implications of negative pressure

Part (c) (Fig. 3) served two purposes: to give students confidence that the negative values obtained in the previous part were appropriate, and to investigate (very minimally) whether students could learn from new ideas provided in the task. More than parts (a) and (b), the design of part (c) was explicitly inspired by studies of the PFL framework. However, due to the constraints of the task and the course comparisons we sought to make (see Section V), we could not incorporate the full design used in seminal studies of the PFL mode of transfer [13,15]. Most notably, we did not allow or assess students' ability to seek out new resources.

c) You should have found different signs for your answers to (a) and (b). In this course, we have not discussed the possibility of negative values of pressure. A more in-depth study of pressure reveals that negative pressures can exist in cohesive substances such as liquids. Just as for positive pressures, a pressure difference across a surface corresponds to a force.

A critical difference between the fluid transport systems of trees and animals like giraffes is that blood vessels through which blood flows are made of a stretchy material, while the xylem through which sap flows are made of a very rigid material.

How do your results for (a) and (b) illustrate part of the reason why trees can grow much taller than land animals? ***Explain your answer using the ideas from this course and your physical intuition. Be as specific as you can be in your explanation.***

**FIG. 3.** Part (c) of the fluids problem. The full task (with additional instructions before the problem, the thermodynamics problem, and the list of equations) is provided in Appendix B.

Our intent was for students to consider the response of an elastic blood vessel (which requires a greater pressure inside than outside to maintain its cylindrical shape and remain open for flow, as the vessel walls are not rigid) vs a rigid xylem (which maintains its shape even if the pressure outside is greater than that inside, allowing it to support a column of water and thereby sustain a negative absolute pressure). We expected this would be very challenging for students, but even so, their responses would give some insight into their ability to learn from and reason with new ideas.

### D. Thermodynamics problem

We included a textbook problem on thermodynamics (Appendix B) to serve as a separate measure of student problem solving skill, as well as to allow the task to serve as a complete "practice test" for students (see discussion of task administration in Section V below), as fluids and thermodynamics had been taught since the last exam. The third part of the problem also involved nontrivial modeling, namely, to recognize that the gas undergoes constant pressure expansion (rather than remaining at constant volume). However, this specific situation had been analyzed explicitly in both the standard and IPLS physics courses.

## V. METHODS: TASK ADMINISTRATION AND ANALYSIS

We administered the task to all students in both courses as long-form problems, and analyzed those students' written responses. Following our analysis of the written responses, we also conducted think-aloud interviews [25,26] with six students in a subsequent offering of the IPLS mechanics course, and analyzed transcripts of those interviews.

### A. Written task administration and responding populations

We wanted students to give the task their best effort, but felt it would be unfair to the non-IPLS students to put the fluids problem on a test, as the standard course had not emphasized modeling choices to the same extent as the IPLS course. We also anticipated that both IPLS and non-IPLS students would find the fluids problem difficult and potentially very stressful. We therefore gave the task to the students as a "practice test" to complete during reading period, between the end of classes and the final exam, with the incentive of getting detailed feedback on their solutions to help them prepare for the final exam. Students were instructed to complete the task under test-like conditions (no use of any resources other than the provided equations and the permitted single sheet of notes) but offered full credit for "completeness and demonstrated effort". Indeed, all submitted solutions displayed significant thought and effort, and there was no evidence of student collaboration or misuse of resources. In both courses, all exams involved long-form problems for which students were required to show work that demonstrated the logic of their solution to obtain full credit, so our "practice test" followed exactly those expectations.

In the IPLS course, 63 of 66 students submitted the problem (two of these 63 students were not life science or pre-medical students and are therefore not included in our study), for a response rate of 95%. In the standard course, over two years, 69 of 128 students submitted the problem. Of those 69, 34 were life science students; the others were engineering students whose responses were not included in our study. In Year 1, the year when a significant number of life science students were enrolled in the standard course, the response rate from the life science students was 47%, very similar to the overall standard course response rate of 53%.

Although the thoroughness of all submitted solutions indicated that all students who completed the task took it seriously, the higher IPLS response rate may indicate that the IPLS students as a group were more invested in completing the task, as the IPLS instructor told the students that he would give feedback, while the standard course instructor told the students that they would get feedback from his physics colleagues. (The study authors provided all feedback for all students in both courses.)

To characterize the population of life science students in the Year 1 standard course who completed the task, we compared overall course grades for those who completed the task and those who did not. Those who completed the task had an average grade of 93.7% ± 0.7% (standard error of the mean); those who did not had an average grade of 90.9 ± 1%. An unpaired *t*-test gives $p$ = 0.02 (two-tailed), indicating that the population who completed the task earned somewhat higher grades than the entire Year 1 life science population. We thus conclude that the life science students from the standard course who completed the task were stronger than average within that course. (We did not do an equivalent analysis for the Year 2 standard course due to the small number of life science students enrolled.) Consequently, any greater success by the IPLS students cannot be attributed to the standard course students being a weaker population overall.

We expect that the overall life science student population in Year 1 should be as strong as that in Year 2, and possibly stronger as students who were less comfortable with or inclined toward physics might delay taking physics until the IPLS course was offered. This therefore suggests that the standard course students may be somewhat stronger on average than IPLS students, and are quite unlikely to be weaker students overall.

### B. Analysis of written work

We developed an emergent coding scheme [26] for student work on the task, based on the types of modeling and problem-solving competencies that were emphasized and cultivated in the IPLS course, as described in section III.B. Two team members (BDG and MT) read through the de-identified, anonymized student work and iteratively developed a code for parts (a) and (b) which documented whether the student's work demonstrated these competencies. Part (c), which gave students new ideas to use in the spirit of preparation for future learning [13], was coded globally for correctness and coherence.

Initially, BDG and MT independently identified the competencies demonstrated by student work, then each developed a rubric for assigning points for competency demonstration. This combination of competencies and rubrics is called the "code." Next, they compared their codes, developed a combined code, independently applied it to a subset of the student responses, and compared results, iterating this process until both coders reached agreement. CHC then applied the code to confirm its reliability (Cohen's kappa was at least 0.85 for all elements). Any remaining disagreements were resolved on a case-by-case basis to generate a final code for analysis, which was applied to all responses by MT. Part of this process involved considering whether the points assigned through the rubric fairly captured the students' demonstrated competencies.

The fluids task code is presented in Table 1, with competencies in the left column and rubric points in the right column; the thermodynamics problem code is presented in Appendix C.

Part (a) (up to 6 rubric points)

| **Competency** | **Rubric** |
|---|---|
| **Model justification** (0-1 pt) | **+1** for showing that $p_2 = p_1 + \rho g \Delta d$ is a special case of the Bernoulli equation<br>**OR**<br>**+1** if the student uses the viscous model (Hagen-Poiseuille) and explains why viscosity is relevant |
| **Reasonableness** (0-1 pt) | **+1** for choosing a heart to brain distance less than 6m |
| **Coordinating equation with physical situation** (0-2 pt) | **+2** for finding that the pressure at the brain is less than the pressure at the heart<br><br>NOTE: Points awarded regardless of whether student uses the fluid statics equation "correctly" (e.g. points awarded even if student assigns a negative value to $g$ or reverses the sign of $\Delta d$ in order to obtain $p_{\text{brain}} < p_{\text{heart}}$) |
| **Coordinating diagram with equation** (0-1 pt) | **+1** for including a diagram that defines $p_1$ and $p_2$ correctly<br>**OR**<br>**+1** for a diagram that labels the heart and brain and clearly demonstrates coordination between the diagram and pressure values in the equation<br>(e.g. labeling the heart and brain in diagram and rewriting hydrostatic equation as $p_{\text{heart}} = p_{\text{brain}} + \rho g(d_{brain}-d_{heart})$<br>**OR**<br>**+0.5** for a diagram that labels the distance between the heart and the brain, regardless of distance chosen (e.g. 6m) |
| **Calculation and numerical skill*** (0-1 pt) | **+1** for correct numerical answer, given the model and height approximation used.<br><br>NOTE: Point awarded for students who find the pressure at the brain is greater than the pressure at the heart, as long as their numerical calculation is otherwise correct. |

Part (b) (up to 6 rubric points)

| | |
|---|---|
| **Model justification** (0-2 pt) | For students who use Hagen-Poiseuille (H-P),<br>**+2** for justifying use of H-P with xylem dimensions (long and skinny),<br>**OR**<br>**+1** for justifying only by stating that the fluid is viscous.<br><br>For students who use Bernoulli,<br>**+2** for attempt to justify using xylem dimensions (even though incorrect)<br>**OR**<br>**+1** for stating that the fluid is non-viscous. |

| | | |
|---|---|---|
| **Flexible coordination of multiple models**<br>(0-2 pt) | For H-P,<br>**+2** for including the gravity term by adding *ρgh* to Δ*p*.<br>**OR**<br>**+1** for considering the role of gravity but not by adding *ρgh* in the calculation.<br><br>For Bernoulli, no points are awardable. | |
| **Model implementation**<br>(0-2 pt) | **Coordinating equation with physical situation** | For H-P,<br>**+1** for correctly implementing $\Delta p = p_{\text{roots}} - p_{\text{leaves}} = \rho g h_{leaves}$<br>**OR**<br>**+0.5** for recognizing Δp is a difference but not properly implementing it.<br><br>For Bernoulli,<br>**+1** if it is used with justification for eliminating velocities (e.g. continuity),<br>**OR**<br>**+0.5** if used correctly with partial justification |
| | **Calculation and numerical skill** | For H-P,<br>**+1** for correct numerical result, given model and value chosen for viscosity<br><br>For Bernoulli,<br>**+1** for correct numerical result given model |

Part (c): up to 2 rubric points, scored holistically for correctness and for coherence/completeness

**Correctness** of reasoning about negative pressure

| **2 pt** | **1 pt** | **0 pt** |
|---|---|---|
| Identifies the **pressure difference across xylem walls** AND notes that the negative sign for xylem indicates a risk of **collapsing inward**. | Identifies pressure difference across xylem walls, but gets the direction wrong. | Does not identify pressure difference across xylem walls. |

**Coherence and completeness**:

| **2 pt** | **1 pt** | **0 pt** |
|---|---|---|
| **Highly coherent and complete**: logically sound explanation for the conclusion. *Clearly and carefully explains all physical mechanisms invoked.* | **Somewhat coherent and complete**: logically sound and mostly internally consistent, addresses the question, but does not explain physical mechanism clearly/carefully | **Not particularly coherent or complete**, or doesn't get at the question asked |

**TABLE I.** Fluids task code (competencies and rubrics), developed as described in text. Task presented in Figs. 1-3.

## C. Statistical analysis

All statistical analyses were performed in the R software environment using appropriate packages. Because Anderson-Darling normality tests indicated that the data were nonparametric, we compared the score distributions from the IPLS and non-IPLS student groups using a Mann-Whitney-Wilcoxon test (also known as a Wilcoxon two-sample test, hereafter referred to as a "Wilcoxon" test), the nonparametric equivalent of a two-sample *t*-test, [27]. (Most of our coded scores had relatively few possible values, as can be seen in Figures 4-7, so this is not surprising.)

Because several of these comparisons found no statistically significant difference between the two groups, we followed with Bayesian analysis to give another measure of the degree of confidence in the null hypothesis, *i.e.*, the equivalence of the two groups. The Bayes factor is the odds ratio of the posterior distributions from a model (the "alternative hypothesis" in statistics terminology) and the null model (or "null hypothesis") [28]. In our study, this corresponds to the odds ratio for a model in which the two groups differ to the null model in which they are equivalent. The Bayes factor can be inverted to give the odds ratio of the null model to the non-null model, thereby giving another measure of confidence in the null model. For example, a Bayes factor of 1:3 comparing the model of differences between the populations to the null model means that the Bayes factor of the inverse comparison is 3:1, indicating three times the likelihood of the null model compared to the alternative. Conventionally, an odds ratio ranging from 1:1 to 1:3 (or the inverse) indicates weak evidence, 1:3 to 1:10 moderate evidence, and 1:10 to 1:30 strong evidence, with even larger odds ratios giving even stronger evidence [28].

For comparisons of frequencies within populations, such as the percentages of different majors or different class years, we used either chi-squared frequency tests, or Fisher's exact test of independence if one or two frequency table entries were small [27].

### D. Think-aloud interview protocol and analysis

We did not administer the written fluids task to entire classes during Year 3 of the study, but recruited a few students to complete the task and participate in a follow-up think-aloud interview. The goal of the interviews was to better understand the reasoning underlying the written responses that had been collected and analyzed in Years 1 and 2; as we were not trying to compare IPLS and non-IPLS students with this work, we did not recruit students from the non-IPLS course. All students in the Year 3 IPLS course ($N = 52$) were invited to participate, and six volunteered.

These six students completed the same written task that had been analyzed in Years 1 and 2, and participated in an approximately 30-minute interview following their completion of the task. Students were not provided any extra course credit for participating, but a small gift card was provided for the time spent completing the written task and doing the interview. The six students who volunteered represented a wide spectrum of student performance in the course: three scored in the top third of the class on the final exam, two in the middle third, and one in the bottom third.

BDG collected electronic copies of the six students' written work on the task and read through them prior to conducting the interviews, which took place within 24 hours of the students having completed the task; students also had access to their written work during the interview. Due to the Covid-19 pandemic, the interviews took place over Zoom. Students were asked to describe their reasoning for each of the three parts of the task, and follow-up questions were asked when clarification was needed. At the end of the interview, students were given an opportunity to ask questions about the task, and to discuss any parts that remained unclear to them.

The interviews were recorded and transcribed. BDG analyzed the six transcripts for general themes and trends, paying particular attention to student reasoning about two concepts: the combining of a gravity term with the Hagen-Poiseuille equation in part (b) of the task, and the meaning of negative pressure in part (c) of the task. Because the only goal in doing the interviews was to develop a better qualitative understanding of the reasoning exhibited in the written tasks that had been carefully analyzed in previous years, and because the interviews were done only with IPLS students, no numerical coding scheme was developed for analyzing the interviews.

## VI. RESULTS AND DISCUSSION

### A. Both groups solve single-model problems with comparable success

Figure 4 compares IPLS and non-IPLS students' rubric scores on parts of the task that involved implementing a simple model that had been previously used in class. Fig. 4(a) shows scores for part (a) of the fluids task – calculating the giraffe's blood pressure at its head. This problem was very similar to others both groups had encountered in their physics courses, although neither course had presented or assigned this problem. A Wilcoxon test (non-parametric test comparing two samples) indicated no significant difference ($p = 0.94$). Likewise, the first two parts of the thermodynamics problem required implementing a model studied in class; Fig. 4b shows the total rubric scores for both groups, again with no significant difference ($p = 0.13$).

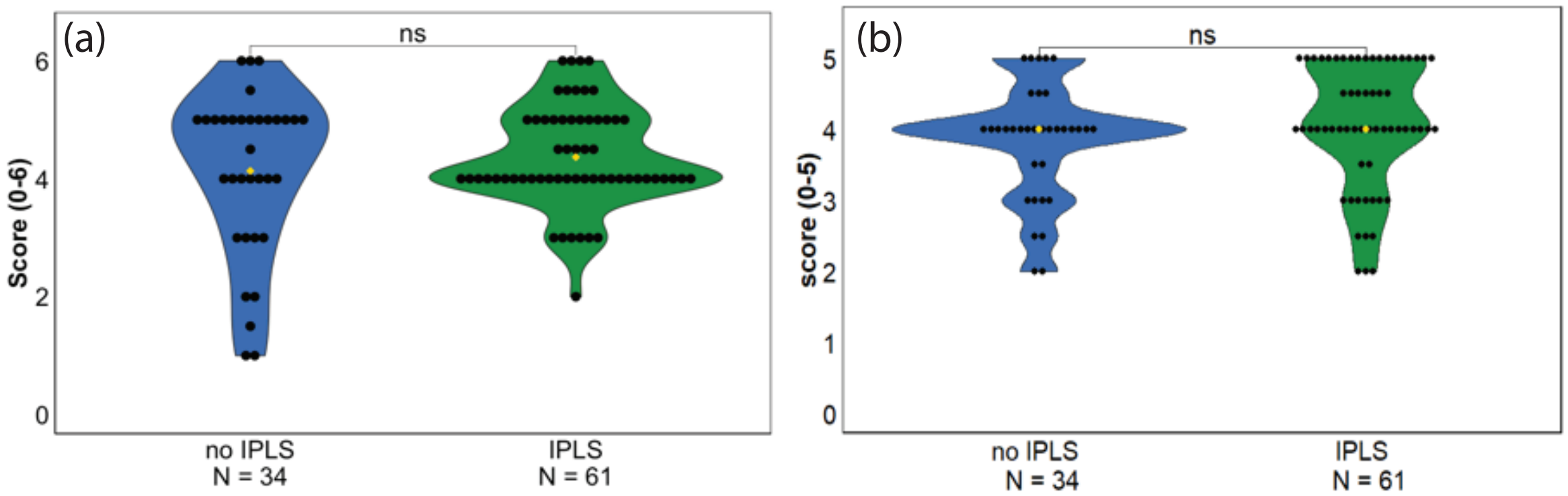

FIG. 4. Total problem-solving scores from non-IPLS students (blue envelope around black dots) and IPLS students (green envelope around black dots) on problems that required implementing a model that had been previously used in class: (a) part (a) of the fluids task (code in Table 1), (b) parts (a) and (b) of the thermodynamics task (code in App. C). Each student's score is represented by a black dot, and the median score is represented by a yellow dot; "ns" stands for not significant and means $p > 0.05$. Bayesian odds ratios in favor of the null hypothesis were (a) 2.8 and (b) 2.3.

For the data shown in Fig. 4, we also carried out Bayesian analysis to determine the confidence with which we could confirm the null hypothesis. For Fig. 4a, we obtained an odds ratio of 2.8 in favor of the null hypothesis (equivalent scores) vs the alternative hypothesis (different); for Fig. 4b, the odds ratio was 2.3. Both correspond to weak confirmation of the null hypothesis. It is likely that the discrete rubric scores with only a small number of possible values limited the effectiveness of modeling the posterior distributions. From this analysis, we conclude that both groups demonstrated comparable skill in solving physics problems that can be solved by application of a single unambiguous model.

We also examined students' basic problem-solving skills across both problems, and found that students from both IPLS and standard courses displayed equivalent levels of skill in carrying out calculations. Fig. 5 displays the total rubric score (up to 5 points) on all code elements labeled "calculation and numerical skill" from both problems (two elements on problem 1 and three on problem 2, each worth 1 rubric point, see Table 1 and Appendix C). The scores are high overall, indicating that students were competent with these basic skills. The distributions are statistically indistinguishable ($p = 0.93$, Wilcoxon test); Bayes analysis gave an odds ratio of 4.4, corresponding to moderate confirmation of the null hypothesis (equivalent skill). We also found that in both groups, the students drew diagrams at indistinguishable rates (roughly half of each group drew them; $p = 0.625$ from a chi-squared test of frequency).

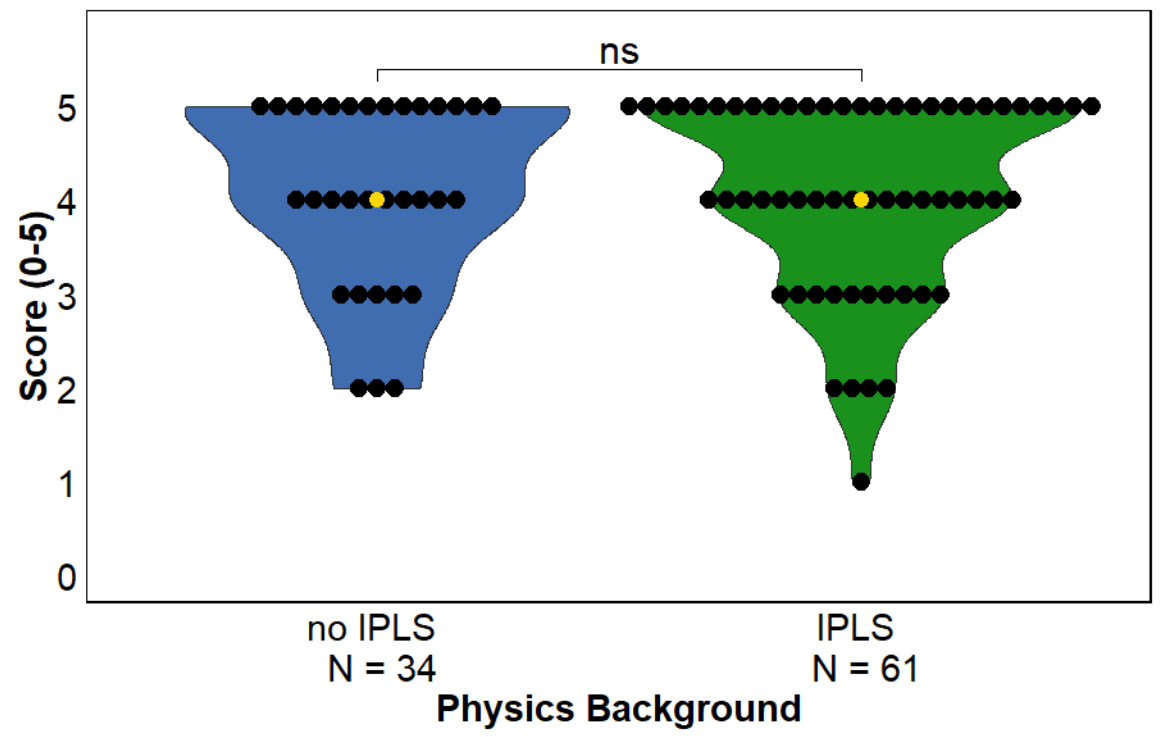

**FIG. 5.** Total calculation rubric scores for non-IPLS students (blue envelope around black dots) and IPLS students (green envelope around black dots). Each student's score is represented by a black dot, and the median score is represented by a yellow dot; "ns" (not significant) means $p > 0.05$.

All of these analyses indicate that IPLS and non-IPLS students in our study achieved similar skill in routine physics problem solving. This is not surprising, as both courses were taught with research-validated methods and both cultivated skills in solving such problems. We see a difference, however, with problems that require more extensive and sophisticated modeling.

### B. IPLS students are significantly more successful with flexible modeling that requires combining concepts

Part (b) of the fluids task requires students to recognize that the fluid flow is viscous based on the dimensions of the xylem. A student who failed to consider the dimensions of the xylem, and searched for an equation that superficially matched information provided in the problem about height and velocity, might choose to use the Bernoulli equation instead. 93% of IPLS students used a viscous flow model (the Hagen-Poiseuille equation), while only 69% of non-IPLS students did, a significant difference ($p = 0.004$, Fisher's exact test). Nearly all students who used the viscous flow model justified it based on the dimensions of the xylem. The much higher rate at which IPLS students modeled the system with viscous flow may have arisen from the greater attention to model choice given in the IPLS course, greater time being devoted to viscous flow in the IPLS course, or both.

A fully correct solution requires not only recognizing that the flow is viscous, but also combining two different models in a way not specifically encountered in the course. While 46% of IPLS students ($N = 28$) combined two models together to solve problem 1b, only a single non-IPLS student did, a difference which is highly significant ($p = 5 \times 10^{-6}$, Fisher's exact test).

Overall, as shown in Fig. 6a, IPLS students were much more successful in solving problem 1b, as indicated by higher overall rubric scores (Δ(median) = 2, $p << 0.0001$). We also found that the IPLS students outperformed the non-IPLS students on the elements of the problem 1b code that are specific to modeling decisions, omitting the points awarded for model implementation, as shown in Fig. 6b. This difference corresponds to an effect size (Cliff's delta) of 0.70 for the total scores and 0.74 for the modeling decision scores.

Part (c) of the thermodynamics problem also required making a modeling choice, but one that had been introduced in class (choosing whether to calculate heat capacity at constant volume or constant pressure). The IPLS students were more successful at modeling this problem (Fig. 6c), with a median score of 2 rather than 1.5, but not significantly so ($p = 0.14$, Bayes factor for the odds ratio of the alternative hypothesis over the null hypothesis of 0.68).

Finally, on both parts (a) and (b) of the fluids task, careful thought was required to apply the equations for pressure correctly to the physical situation. Combining the scores for the "coordinating equation with physical situation" codes from 1a and

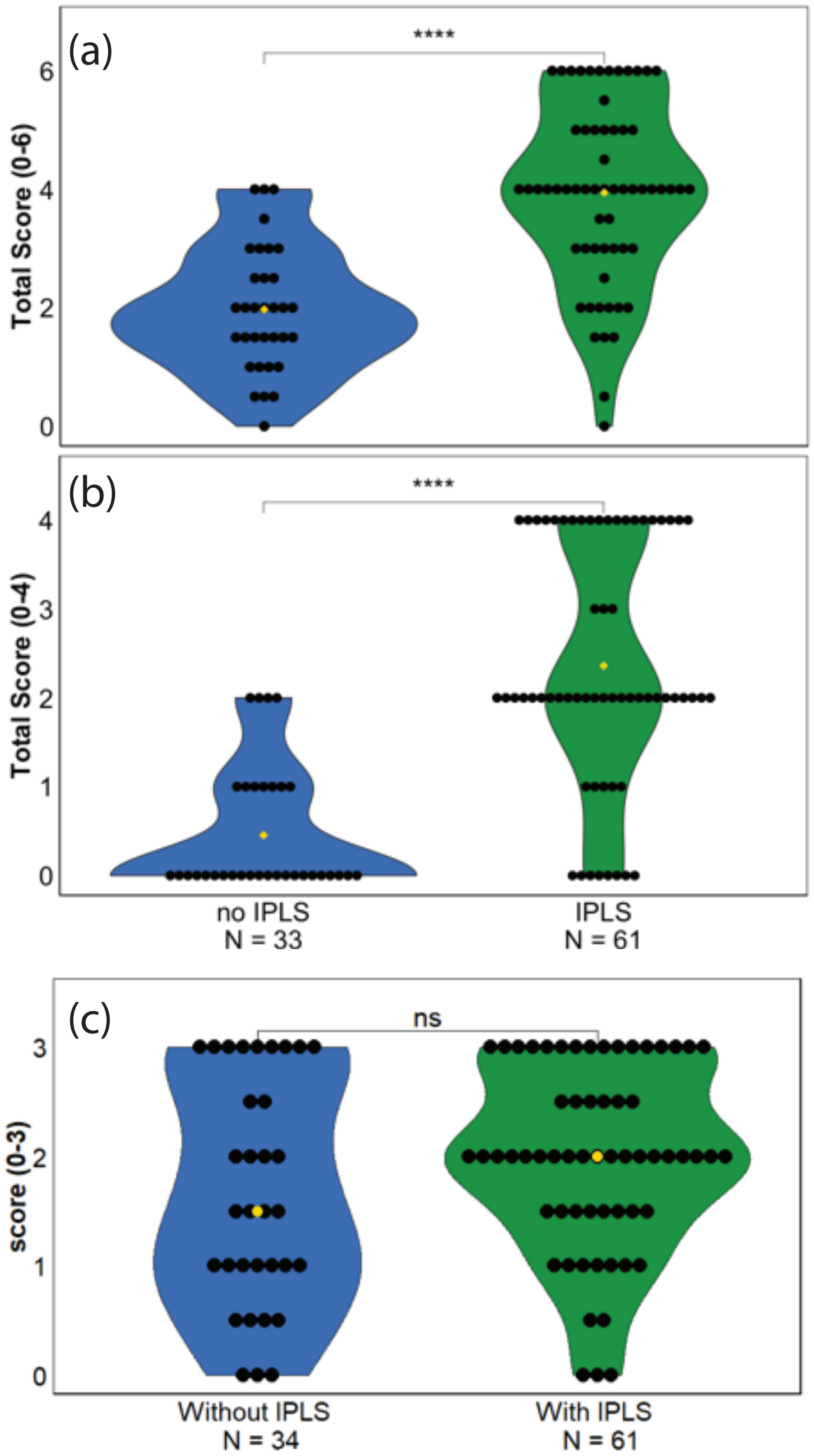

FIG. 6. Non-IPLS student scores (blue envelope, black dots) and IPLS student scores (green envelope, black dots) on part (b) of the fluids task: (a) total score over all elements, (b) score on just the justification and modeling parts of the task; (c) scores on part (c) of the thermodynamics task, which required some modeling. One non-IPLS student skipped part (b) of the task but completed other parts, giving different $N$ for (a),(b) than for (c) and Figs. 4, 5. Each student score is represented by a black dot, and the median score is represented by a yellow dot; **** indicates $p < 0.0001$, "ns" (not significant) means $p > 0.05$.

1b together, the IPLS students (median score 3.5) were significantly more successful than the non-IPLS students (median score 2.5; $p$ = 1x10$^{-4}$, Wilcoxon two-sample), as shown in Fig. 7, suggesting that the IPLS course may have cultivated attention to the physical meaning of equations in a manner that the standard course did not.

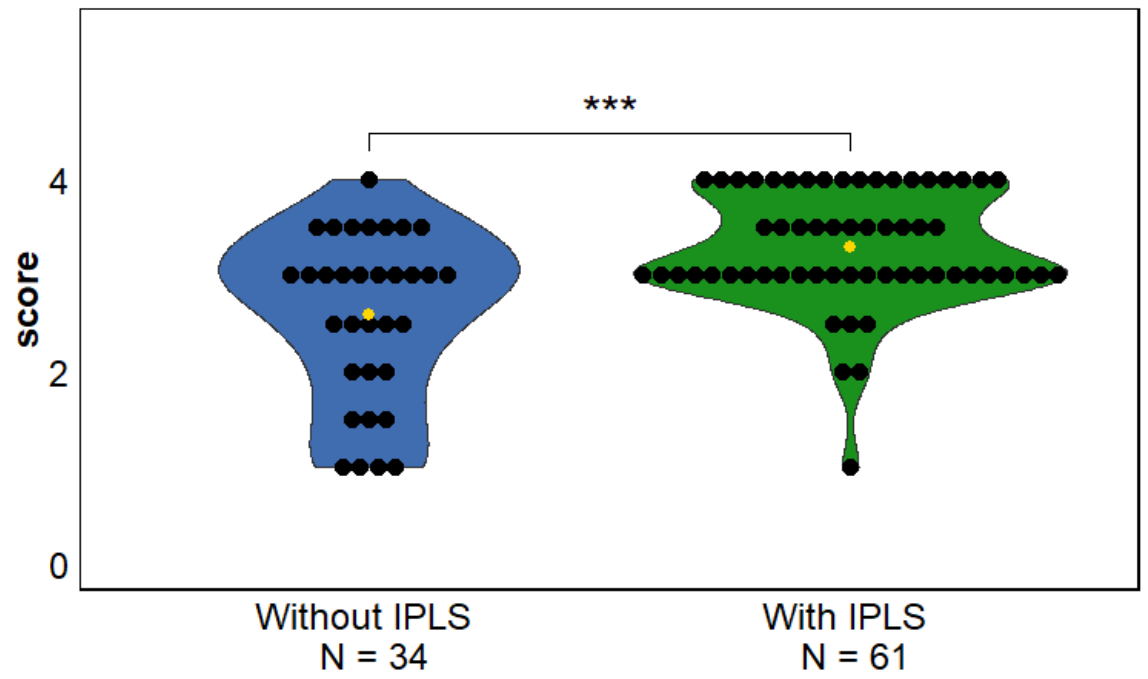

**FIG. 7.** Overall rubric scores for coordinating equations to physical situations from non-IPLS students (blue envelope, black dots) and IPLS students (green envelope, black dots) from parts (a) and (b) of the fluids task. Each student's score is represented by a black dot, and the median score is represented by a yellow dot. *** indicates $p < 0.001$.

While it is beyond the scope of this study to determine why the IPLS students were more successful at such sophisticated modeling tasks, we expect it is due to the emphasis in the IPLS course on the process of modeling, implemented through problems which explicitly require students to make modeling choices, as described in section III.B. If so, this result validates the importance of those pedagogical choices to develop modeling competencies in students.

### C. IPLS students provide more coherent discussions when using new knowledge

On part (c) of the fluids task, both IPLS and non-IPLS students struggled to articulate mechanistic and physically accurate explanations for why trees (but not animals) can withstand the negative pressure calculated in the previous part of the task. Many students from both groups brought in specific biological knowledge to try to account for the negative result, but few did so in a way that clearly demonstrated complete, correct thinking about the physical mechanism.

The large negative pressure inside the top of the xylem obtained in the previous part of the problem corresponds to a radially directed inward force, as the space in the tree trunk surrounding the xylem is at atmospheric pressure. Stiff xylem, but not flexible blood vessels, can withstand such an inward force. Only about 40% of the students in the study even mentioned this idea of inward-outward (radial) force, with the remaining 60% providing explanations that involved only vertical differences in pressure (associated with height differences). This percentage was about the same for both groups. Furthermore, among the approximately 40% of students who did mention the inward-outward force, only about one-third of these students correctly identified the direction of the force as inward. Many students did not specify a direction, and some incorrectly thought the net force would be outward. Here, too, the results were similar across both groups.

While students from both the IPLS and standard courses found it similarly challenging to provide a fully correct physical mechanism underlying the negative pressure result (as measured by a "correctness" score in our code for 1c, Table I), the IPLS student responses to this part of the task were more *coherent* (as measured by a "coherence" score in our code for 1c, Table I) with $p = 0.0002$ (chi-squared test of frequency, $\chi = 17.6$). Specifically, for coherence, 18 of the 60 IPLS students (30%) received 2 rubric points, 40 received 1, and only 2 received 0 (and one more student wrote nothing), while only 6 of the 34 non-IPLS students (18%) received 2 rubric points, 16 received 1, and 12 received 0. The higher coherence scores from IPLS students could reflect the repeated emphasis on providing clear written explanations throughout the IPLS course.

A response received a higher score on the coherence dimension when it addressed the question that was asked, and was both internally consistent and logically sound. That is, a student's response was judged to be highly coherent when the conclusions followed logically from the specific physical principles stated, and when it did not include extraneous physical or biological principles that were unrelated to the conclusions drawn.

To illustrate what makes a response coherent, one response that received full points for coherence included the statement "a negative pressure means the force is outward, and because the xylem in the tree are stiffer than the blood vessels, the tree can withstand a greater outward force without exploding." Although this response incorrectly associates a negative pressure with an outward explosion (and therefore did not receive a full score for correctness), the student relates the sign of the pressure to the direction of a force across the vessel walls, and relates the ability to withstand this force to the stiffness of the vessels. This is an internally coherent response that relies on the relevant physical mechanism and does not bring in irrelevant considerations. A response that scored lower for coherence attributed the large negative pressure at the top of the tree to the "need for nutrients to reach much greater heights in trees than they do in giraffes." This response is redundant (the negative pressure is due to the tree's significant height, as already calculated earlier in the problelm) and brings in biological ideas about energy consumption that are neither relevant to this particular question nor mechanicstic. It is not incorrect that the tree requires nutrients to reach greater heights than does the giraffe, but this does not explain *why* the tree is able to withstand the large negative pressure.

### D. Think-aloud interviews

In their written responses to part (b) of the fluids task, all six students who participated in the think-aloud interviews correctly chose a viscous flow model for analyzing sap flow through the xylem, and all six students cited the long and narrow dimensions of the xylem tubes as the key indication that they must consider viscous flow. It was clear from the interviews that these students had developed a firm association between the dimensions of a tube and the appropriateness of a particular physical model for fluid flow.

In their written work, three of the six students who participated in the think-aloud interviews also successfully combined a gravity term with the Hagen-Poiseuille model, comparable to the 46% of IPLS students who did this successfully in the written responses from the previous year. When asked to elaborate on their reasoning during the interviews, two of these three students described how the pressure difference between the bottom and top of the tree needed to be sufficiently great so as to overcome *both* viscous resistance (as represented by the Hagen-Poiseuille equation) and gravity (as represented by the hydrostatic equation from part (a) of the task). The third student who solved part (b) correctly did not articulate their reasoning with the same clarity, but conveyed a similar idea, saying that "the Hagen-Poiseuille part was for horizontal, but there also has to be a height component for a tree."

When the three students who did not successfully combine the gravity term with the Hagen-Poiseuille equation were asked to "explain their reasoning," none were initially bothered by the absence of a gravity term. When prompted by the interviewer to consider "whether it matters that the tree is vertical rather than horizontal," all three agreed that it should matter, but struggled to articulate how they would account for it. None of these three students reached a resolution on their own during the interview.

The six students' written responses to part (c) of the task varied in correctness and coherence, as was true for the written responses analyzed the prior year. Three of the six written responses did not address the question explicitly asked — why negative pressures could be sustained by a tree (and by its rigid xylem) but not a giraffe (and its elastic blood vessels)— and simply stated that the negative pressure was due to the tree being taller than the giraffe (when the height difference is larger, the pressure difference between top and bottom is larger, leading to a negative number at the top). Four of the six interviewed students spoke at length about various *biological* aspects of the scenario when explaining their responses to part (c) of the task, implicitly associating the negative pressure with some "need" that the tree had to supply energy or nutrients to great heights. Two of these students said that the negative pressure was important for assuring greater "energy efficiency" and the other two students suggested that the negative pressure was somehow important for transpiration at the leaves.

These biological details were salient to the students, even though they were ultimately irrelevant for understanding *why* xylem can withstand negative pressure. This challenge of clearly distinguishing physical mechanism from biological function when providing explanations in the IPLS context has been identified elsewhere and was therefore not unexpected [29]. Some of the interviewed students explicitly described viewing the prompt in part (c) of the fluids task as inviting a biological explanation, rather than a physical mechanism. As one of them put it, "I thought we were supposed to give a more biological explanation for that part."

Just one of the six students who participated in the think-aloud interviews spontaneously provided a fully correct description of the significance of the negative pressure in part (c) of the task. That student described the xylem stiffness as preventing an inward-directed collapse due to the negative pressure, and used hand motions to indicate the direction of the forces involved. This student also spontaneously described the difference between the situation in the task and one involving positive pressure, using the case of an ideal gas pushing outward on the walls of a container as an example. It was clear from this response that the student was interpreting negative pressure as being related to an inward directed force on the xylem. A second student described the negative pressure as being associated with radially-directed forces, but incorrectly stated that the tree was in danger of exploding outward rather than collapsing inward. The remaining students who were interviewed struggled to spontaneously articulate a mechanistic account for why trees (but not giraffes) could withstand negative pressure. However, when prompted by the interviewer to consider radially directed forces (rather than vertical pressure differences), every one of the six students was able to provide a coherent account of why the xylem stiffness would be relevant and important.

The think-aloud interviews crystalized two aspects of IPLS student reasoning that were suggested by the written responses collected the previous year:

- Students who successfully included a gravity term in their written response to part (b) of the task likely did so because they recognized that the Hagen-Poiseuille model for viscous flow (implicitly) assumed a horizontal tube, and did not account for gravitational resistance to fluid flow. These students were able to flexibly modify the Hagen-Poiseuille model to account for this complication, even though they had never before been asked to do this. Although this study has not examined how students developed these abilities, we believe this is most likely developed through the kinds of problem solving tasks used in the course as described in section III.B.
- The vague nature of student responses to part (c) of the written task may reflect not only the struggle to understand the physical significance of negative pressure (i.e., that it is related to radially directed force), but also life science students' likely interpretation that part (c) was an invitation to elaborate on their biological understanding of the nutrient transport system in trees. IPLS students were more likely than non-IPLS students to provide *coherent* (internally consistent and logically sound) explanations for the negative pressure in their written responses, perhaps because of the frequency with which they had been asked to provide thorough reasoning during the semester, but were not more likely to provide fully *correct* physical mechanisms.

## VII. CONCLUSIONS AND OUTLOOK

This article presents our findings on IPLS and non-IPLS students' ability to carry out a sophisticated biological modeling task at the end of first-semester introductory physics. We found that the IPLS students were dramatically more successful at building a model which combines multiple ideas they had not previously seen combined, and at making complex decisions about how to apply an equation to a particular physical situation. This seems unlikely to correspond to differences in basic problem solving ability or calculational skill; both groups of students carried out calculations, and identified and applied simple models that they were introduced to in class, at statistically indistinguishable rates. Rather, it seems likely that the difference is due to some aspect of the IPLS course. As our analysis is correlational rather than causal, we cannot rule out the possibility of another confounding factor.

In addition to these results, the task and the reasoning that went into its development are also contributions. We consider this task to be a suitable instrument for assessing modeling skills, and welcome others to use it. We hope our description of the development guides other investigators to develop tasks around other physics content.

Our study does not directly investigate *how* the IPLS course might have accomplished this difference, although we describe in Section II the key features of the course that we believe to be relevant. In particular, the IPLS course explicitly emphasized modeling skills, particularly in the fluid dynamics unit. The Bernoulli equation and the Hagen-Poiseuille equation were introduced as non-viscous and viscous models of fluid flow, respectively, and students solved fluid dynamics problems both in class and on homework that explicitly required them to choose between these two models. The IPLS course also strongly emphasized that many equations apply in only limited cases. (The textbook used for both courses also emphasizes this, so the non-IPLS students had the opportunity to learn this as well. The in-class discussion of the Hagen-Poiseuille equation in the standard course may have been less emphatic that the equation applies only to horizontal pipes.)

In a companion study, our research team found that the IPLS course cultivated durable increases in students' perception of the value and relevance of physics for biology [30]. This could have led to the greater success of the IPLS students in a variety of ways. Although as equivalent groups of life science majors, both groups would presumably have been equally likely to find the flow of sap through trees interesting (or not), the problem is fairly technical and the analysis required is quite challenging. Success required a combination of creativity and courage that may have been more likely in a group that had come to see physics as relevant and valuable for understanding biological phenomena. Finally, IPLS students may have been more motivated in approaching physics course tasks than non-IPLS students, due to experiencing the course as a whole as valuable, which could also have led to improved performance.

Further study is needed to identify which of these possible mechanisms are at work, and whether there are others. It is also yet to be determined whether IPLS students are more successful at complex modeling in general or solely in biological settings, and whether this skill is durable.

In work reported separately, we found that former IPLS students performed more strongly than those who had not taken IPLS on a task that required them to quantitatively and mechanistically analyze diffusion in a biological problem presented in a biology capstone course [12]. That work did not, however, explore students' ability to employ models in a flexible way.

## ACKNOWLEDGMENTS

We gratefully acknowledge Eugenia Etkina for suggesting a transfer task at the end of introductory physics, and Wing-Ho Ko for partnership in designing and offering a task that could be used with both IPLS and standard introductory mechanics. We thank the advisory board for our longitudinal study (Todd Cooke, Eric Brewe, and Eric Kuo), and many colleagues, including Sara Hiebert Burch, Chandra Turpen, and Joe Redish, for valuable discussions. This work was funded by the National Science Foundation under the IUSE program (EHR-1710875).

**APPENDICES**.

APPENDIX A. Problem sequence applying fluid dynamics to cardiology, used in IPLS course.[1]

A. Your circulatory system includes many vessels and valves with different cross-sectional area. Blood has relatively low but not completely negligible viscosity.
   - In which vessels is the flow best described with viscous flow?
   - In which is non-viscous flow the best description?[2]

B. In aortic stenosis, the open area of the aortic valve has changed from its normal value. The patient's heart adjusts to maintain the same flow rate as in a healthy individual, until it can no longer change. Based on the flow velocity data shown, how does the valve area of the patient with aortic stenosis, $A_{stenotic}$, compare to the valve area of the normal person $A_{normal}$?

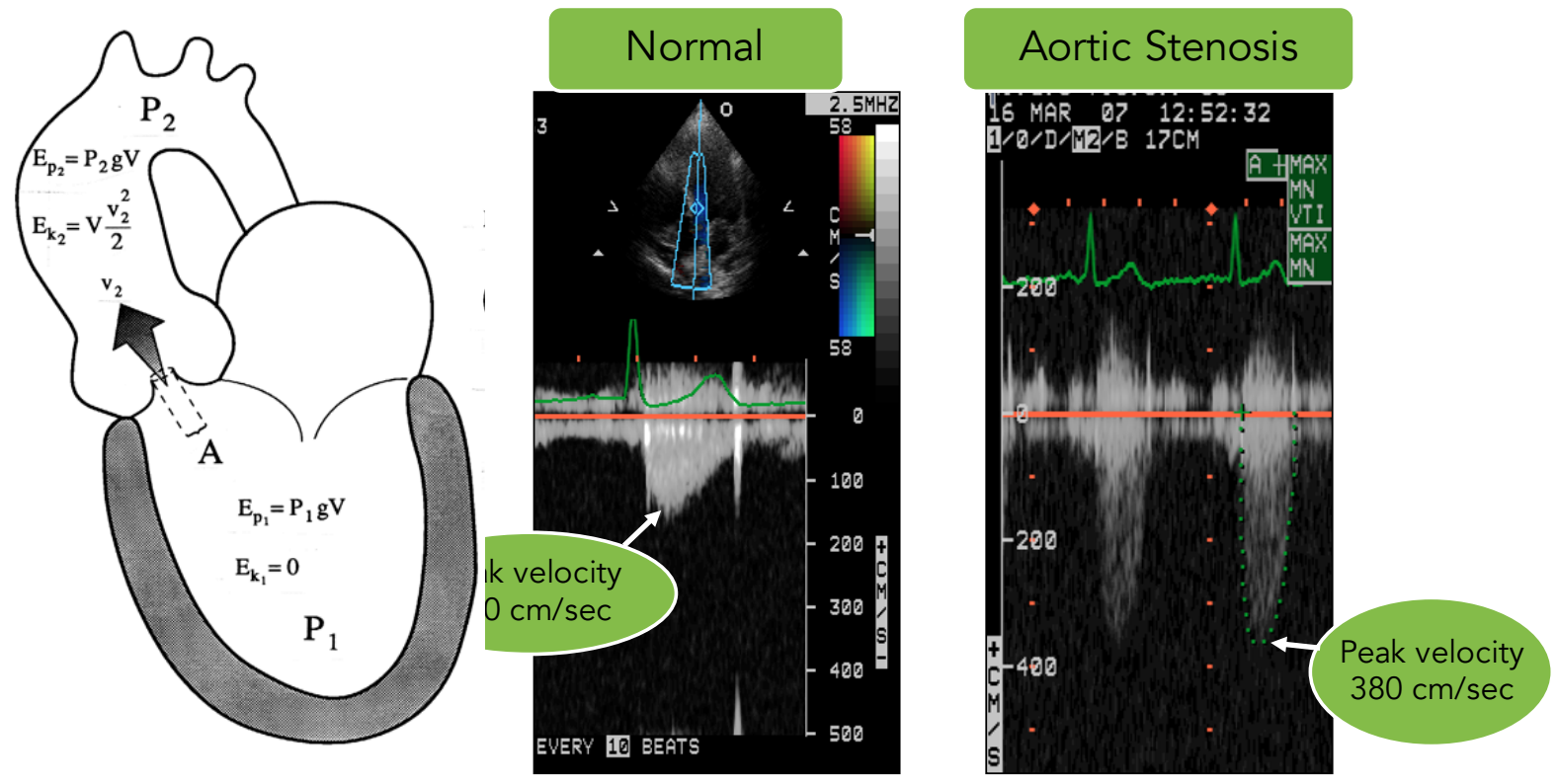

C. Based on the figure to the right, which model is most appropriate for blood being pumped through the aortic valve, the non-viscous model or the viscous model?[3]

*After answering, discuss: Why does the model you chose make sense (i.e. if you chose the viscous model, why would you expect there to be a lot of viscous interactions? If you chose the non-viscous model, why would viscous interactions be minimal? Look at the equation for the pressure drop in the non-viscous model.)*

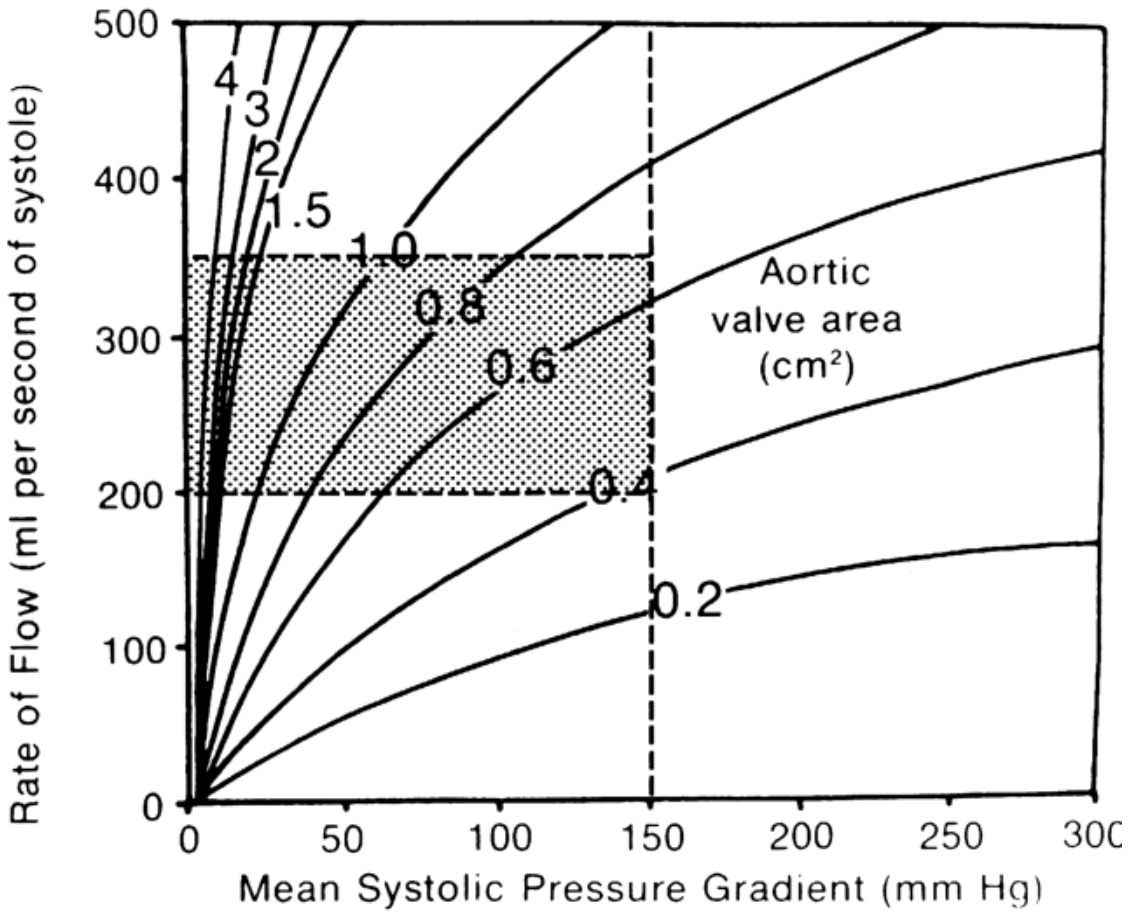

D. Which branch gets the largest fraction of the total flow?

*If you finish: The circulatory system can adjust the amount of flow that goes to each branch dynamically. How might it do that? What feature of the viscous flow equation makes that possible?*

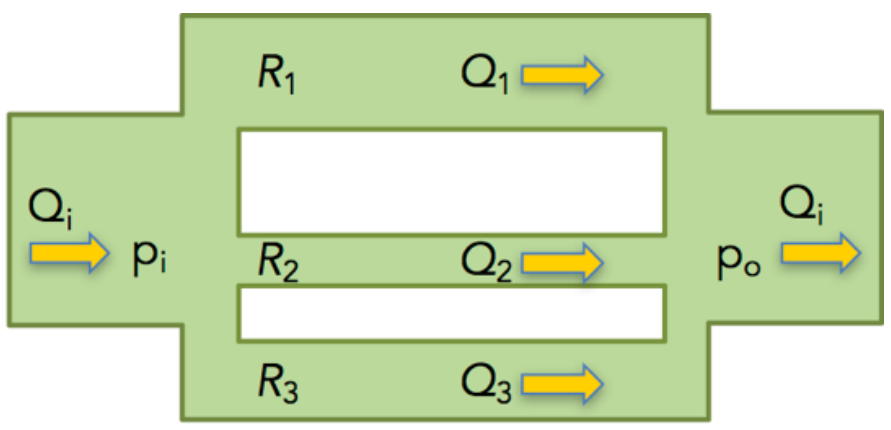

---

[1] Developed by CHC, BDG, and John W. Hirshfeld Jr; available through livingphysicsportal.org, CC-BY-NC-ND 4.0
[2] Hint: Think about where the frictional losses happen. Would these be greater for small or large diameter vessels?
[3] Hint: Look at the axes of the graph and figure out what is shown. What kind of relationship between pressure difference and flow is predicted for viscous flow? For non-viscous flow? Which matches the graph?

APPENDIX B. Complete task, with instructions and equations provided to students.

*Note to instructors/researchers: If students estimate the height of the giraffe's heart to be 3 m or less, and estimate the brain to be at 6 m, they will find the pressure in the brain is below atmosphere. As the pressure is most easily calculated in Pa, all of our students found the value in Pa and almost none noted that it was below atmospheric pressure. If in using this task you wish to avoid this possible complexity, you could make modifications, such as eliminating the estimation from the photo and just give the students the height of the heart as 3.5 m.*

**Practice problems for feedback, due to homework box by Monday December 17 at 10 AM**
***Turning in your solutions by Friday 6 PM will get you feedback by Saturday afternoon!***
**Worth 1/2 homework assignment, full credit for completeness and demonstrated effort**

*These problems bring together multiple ideas from the course, particularly the material since the second midterm (fluids and thermodynamics). They are provided for you to practice solving problems in a test-like situation and receive feedback before the final exam. Please solve these problems by yourself (no books or talking to fellow students; feel free to use your page of notes if you have it prepared).*

*Take as much time as you like; the problems are designed so that you should be able to complete them in up to 45 minutes if you are well prepared. If you give these problems your best effort, it will give you the most useful feedback!*

*The last page gives equations and values of useful parameters such as the density and viscosity of water.*

Problem 1.

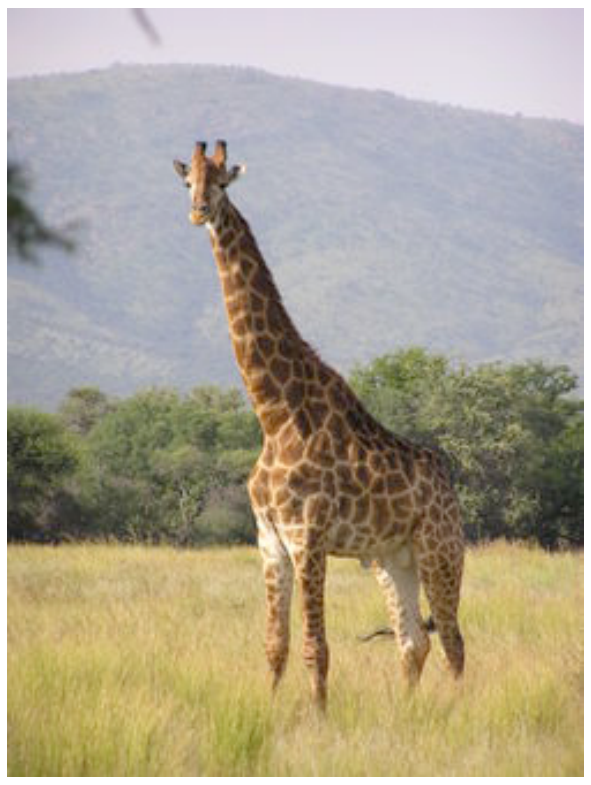

(a) Adult male giraffes can reach a height of roughly 6 m. The minimum pressure of the blood[4] leaving the giraffe's heart is 1.24 atmospheres (124 kPa). Find an approximate value for the minimum blood pressure in the giraffe's brain when its neck is extended to its full height. You may infer information from the picture of a giraffe[5] at the right.

***Please briefly explain the reasoning you used to find your answer, including how you decided which equations to use, as well any approximations you made. Also please show your work.***

(b) In trees, water is carried from the roots to the leaves by the flow of sap (water with other kinds of molecules dissolved in it) through stiff tube-like structures, called xylem. Although sizes vary, a typical diameter would be 100 μm. In the main trunk of the tree, they extend close to the full height of the tree, which is commonly as great as 30 meters tall or taller (5 species of tree are known to reach 90 -110 m in height). These extremely narrow, long tubes, called xylem, contain a continuous column of water which can then flow into the leaves. The evaporation of water from the leaves (called transpiration) causes water to be steadily drawn into the leaves from the xylem. The structure of the leaves allows the pressure of water in the xylem to not necessarily be the same as the surrounding atmospheric pressure.

[4] Although blood is a mixture of water and various types of blood cells, the density of blood is very close to the density of water because the cells also consist mostly of water.
[5] Wikimedia Commons, Miroslav Ducacheck, CC BY-SA 3.0.

Consider a tree in which sap flows through each 100 µm-diameter xylem at a volume flow rate of $1.1 \times 10^{-10}$ $m^3/s$ (equal to $1.1 \times 10^{-4}$ mL/s or 0.40 mL/hr), corresponding to an average flow speed of 0.014 m/s. (Given the huge number of xylem, the total flow for the entire tree is substantial!) If the pressure in the roots is equal to atmospheric pressure, what is the pressure at the top of a 30 m tall xylem in the trunk?

***Please briefly explain the reasoning you used to find your answer, including how you decided which equations to use, as well any approximations you made. Also please show your work.***

**(c)** You should have found different signs for your answers to (a) and (b). In this course, we have not discussed the possibility of negative values of pressure. A more in-depth study of pressure reveals that negative pressures can exist in cohesive substances such as liquids. Just as for positive pressures, a pressure difference across a surface corresponds to a force.

A critical difference between the fluid transport systems of trees and animals like giraffes is that blood vessels through which blood flows are made of a stretchy material, while the xylem through which sap flows are made of a very rigid material.

How do your results for (a) and (b) illustrate part of the reason why trees can grow much taller than land animals? ***Explain your answer using the ideas from this course and your physical intuition. Be as specific as you can be in your explanation.***

Problem 2.[6]

A 6.0-cm-diameter cylinder of helium (He) gas has a 1 kg movable copper piston. The cylinder is oriented vertically, as shown,[7] and the air above the piston is evacuated. When the gas temperature is $20^0$C, the piston floats 20 cm above the bottom of the cylinder.

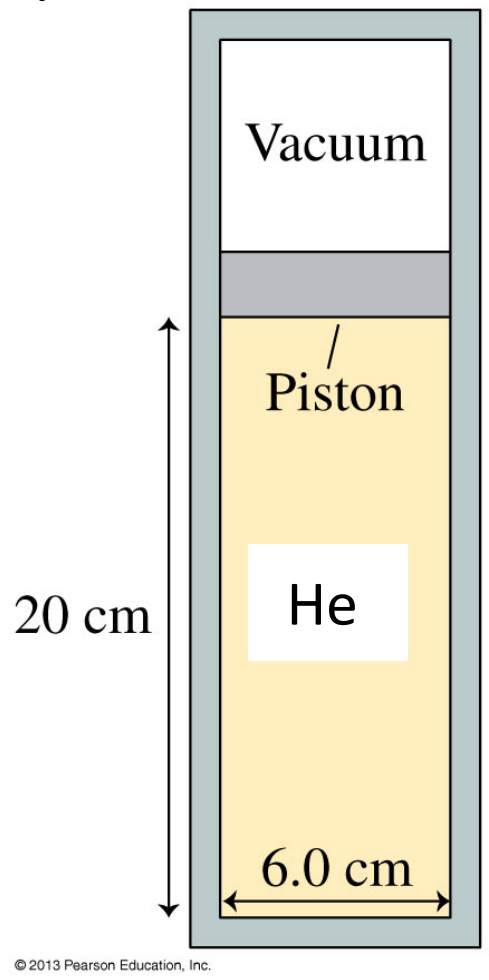

In answering the following questions, be sure to ***explain your reasoning***, and ***explain why you choose to use the equations that you use.***

a) What is the gas pressure?
b) How many moles of gas are in the cylinder?

Then 2.0 J of heat energy are transferred to the gas.

c) What is the new equilibrium temperature of the gas?

[6] Adapted significantly from Randall D. Knight, *Physics for Scientists and Engineers*, 3rd edition (Pearson, 2013), problem 17.51. This edition of Knight was the textbook for both courses.
[7] Figure P17.51 from Knight, *Physics for Scientists and Engineers*, 3rd edition (Pearson, 2013).

**Equations for Practice Problems**

Properties of water (at 20°C)

Density: $\rho_{water} = 997\ \text{kg/m}^3$

Viscosity: $\mu_{water} = 1.005 \times 10^{-3}$ Pa • s

Units and constants:
$g = 9.80\ \text{m/s}^2$

$1\ \text{N} = 1\ \text{kg/m} \bullet \text{s}^2$
$p_{atmos} = 1.00\ \text{atm} = 1.013 \times 10^5\ \text{Pa} = 101.3\ \text{kPa}$

$1\ \text{Pa} = 1\ \text{N/m}^2$
$R = 8.31$ J/mol K

$1\ \text{L} = 1 \times 10^{-3}\ \text{m}^3$
$k_B = 1.38 \times 10^{-23}$ J/K

$T_K = T_C + 273$
$N_A = 6.023 \times 10^{23}$ per mole

Atomic mass unit: $1\ \text{u} = 1.66 \times 10^{-27}$ kg

Force of gravity near Earth's surface: $F = mg$

Density, pressure, and hydrostatics:

$$\rho = \frac{m}{V} \qquad F = pA \qquad p_2 = p_1 + \rho g \Delta d$$

Fluid dynamics:

Continuity: $$Q_{volume} = \frac{d(\text{volume})}{dt} = vA$$

Bernoulli: $$p_1 + \rho g y_1 + \frac{1}{2}\rho v_1^2 = p_2 + \rho g y_2 + \frac{1}{2}\rho v_2^2$$

Hagen-Poiseuille: $$Q = \frac{\Delta p}{\ell}\frac{\pi R^4}{8\mu}; \quad v_{avg} = \frac{\Delta p}{\ell}\frac{R^2}{8\mu}$$

First law of thermodynamics: $\Delta E_{th} = W + Q$

Specific heat and calorimetry: $Q = Mc\Delta T = nC\Delta T$

Ideal gas properties:

$$pV = Nk_B T \text{ and } pV = nRT$$

$$\varepsilon_{avg} = \frac{1}{2} m_{atom} v_{rms}^2 = \frac{3}{2} k_B T$$

$$K_{micro} = \frac{3}{2} pV = \frac{3}{2} N k_B T$$

$$C_P = C_V + R$$

For a monatomic ideal gas, $E_{th} = K_{micro}$ and $C_V = \frac{3}{2}R$

Work done in ideal gas process $$W = \int_{V_i}^{V_f} p\,dV$$

APPENDIX C. Solution to task (provided on paper to students along with feedback).

**Problem 1.**

**1a.** This pressure difference is calculated just using static pressure, $p_2 = p_1 + \rho g \Delta d$ ; because blood velocities are slow, although you could try using the Bernoulli equation for nonviscous flow, it turns out that the flow terms are very small compared to the gravity term. (You also might have figured you could only use static pressure because you aren't given enough information to calculate anything due to flow. If you noticed this but wondered why it is reasonable that the flow rates could be neglected, good job!)

Estimating that the giraffe's heart is a bit higher than halfway up its body, so that the distance from the heart to the brain is about 2.5 m, gives pressure at the brain = 100 kPa. If you estimated a slightly different distance, that's fine. Using 6 m for the distance assumes the heart is at the bottom of the feet which is not reasonable.

**1b.** Flow in the xylem is viscous, which we know because the tubes are so narrow and long. Consequently there is a pressure difference from bottom to top given by the Hagen-Poiseuille equation:

$$Q = \frac{\Delta p}{\ell}\frac{\pi R^4}{8\mu}; \quad v_{avg} = \frac{\Delta p}{\ell}\frac{R^2}{8\mu}$$

The Hagen-Poiseuille equation is derived for horizontal flow. To account for the effect of gravity, we add the hydrostatic pressure difference calculated as in (a).

The viscous term gives a pressure difference of $\Delta p = \frac{8\mu Q\ell}{\pi R^4} = 1340\text{kPa}$ (note that $Q$ is the volume flow rate not the flow speed; if you calculate the pressure difference from the flow speed you need to use the second version of the equation above) and the gravity term gives $p_{bottom} - p_{top} = \rho g \Delta d = 293\,\text{kPa}$.

The viscous flow term also makes the pressure at the top lower than at the bottom, because the fluid is flowing from bottom to top.

Combining them gives $p_{top} = p_{bottom} - \Delta p = -1532\,\text{kPa}$

**1c.** Negative pressure inside the xylem while the pressure outside is positive atmospheric pressure means means there is an enormous inward force on the xylem walls. The xylem walls are rigid so they can withstand this. However, a stretchy blood vessel in an animal can't withstand it and would collapse. Thus blood pressure in animals has to always be positive.

**Problem 2.**

**2a.** The piston is stationary, so the force exerted by the gas pressure upward on piston must balance the weight of the piston down, giving $p_{gas}$ = 3.5 kPa. The area of the piston on which the gas pushes is the area of the circular base, with radius 3.0 cm = 0.030 m. Express all lengths in meters to make the units work out correctly.

**2b**. We can model He as an ideal gas, and we want the number of moles *n*, so use the ideal gas law in the form $pV = nRT$ with the dimensions in the problem and the pressure calculated in (a). This gives $n = 8.0 \times 10^{-4}$ mol. ($R$ is given in units of J/mol K; this works if $p$ is in Pa, $V$ is in m$^3$, and $T$ is in K.)

**2c**. Start by determining whether this is a constant pressure or constant volume process. Pressure is kept constant by the piston's weight. Therefore ΔT is determined from $Q = nC\Delta T$ (where $n$ is the number of moles) by using the molar heat capacity at constant pressure, which for a monatomic ideal gas is $C_P = C_V + R = \frac{3}{2}R + R = \frac{5}{2}R$. This gives the final temperature as 140 °C = 413 K.

APPENDIX D. Code for thermodynamics problem.

**Part A (up to 3 points)**

| **Competency** | **Scoring Criteria** |
|---|---|
| Using an appropriate model (0-1 pt) | +1 for using $p = F/A = mg/A$ to find pressure. Award points if the student writes $F = ma$ but substitutes gravity for acceleration.<br>OR<br>+0.5 for using $p = F/A = mg/A$ but student equates area to surface area of cylinder or width•diameter of figure. |
| Coordinating a diagram with an equation (0-1 pt) | +1 for correct FBD of piston given model chosen,<br>OR<br>+0.5 for FBD with incorrect or incomplete labeling of the forces. |
| Numerical calculation and facility with units* (0-1 pt) | +1 for solving correctly for $p$ (including units), given the equation used. |

**Part b (up to 2 points)**

| **Competency** | **Scoring Criteria** |
|---|---|
| Using an appropriate model (0-1 pt) | +1 for using $pV = nRT$. |
| Numerical calculation and facility with units (0-1 pt) | +1 for solving correctly for $n$ (including units), given the equation used. |

**Part c (up to 3 points)**

| **Competency** | **Scoring Criteria** |
|---|---|
| Using an appropriate model (0-1 pt) | +1 for using the appropriate model: $Q = nC_p\Delta T$ or $Q = (5/2)nR\Delta T$<br>OR<br>+1 for deriving the appropriate model:<br>$Q =\Delta E - W$, with $\Delta E = (3/2)nR\Delta T$ and $W = -p\Delta V = -nR\Delta T$.<br>OR<br>+0.5 for a model involving $Q = nC\Delta T$ without recognizing constant pressure. This can include:<br>• $Q = nC_V\Delta T$<br>• $Q = mc\Delta T$, where m=n*molar mass<br>• $\Delta E=(f/2)NR\Delta T$, with first law justification |
| Model justification (0-1 pt) | +1 for a justification of choice between $C_V$ and $C_P$.<br>Stating that $C_V$ OR $C_P$ applies (without further justification) is sufficient; students only need to recognize that different heat capacities apply under different conditions.<br>OR<br>+1 if an incorrect model is used but a rigorous justification for the model choice is provided. |
| Implementation and numerical calculation and facility with units (0-1 pt) | +1 for solving correctly for $T$ (including units), given the equation used, including finding $C_V$ or $C_P$ correctly. |

***Numerical calculation and facility with units competency clarification:**

| Award point | Do not award point for... |
|---|---|
| "Careless" or trivial substitution error<br>• Ex. substitute 1.1 x $10^{-2}$ instead of 1.1 x $10^{-3}$<br><br>Award point regardless of the equation copied or utilized as long as the following computation follows | Clear algebraic error<br>• Ex. simplify $\pi R^4$ to $A^2$<br><br>Meaningful substitution error<br>• Ex. substitute diameter for R<br><br>Unit conversion and magnitude errors |